\documentstyle[12pt]{article}

\begin{document}
\def\cd{\cdot}
\def\qth{q_{\theta}}
\def\thv{\vec{\theta}}
\def\yv{\vec{y} }
\def\gh{\hat{\gamma} }
\def\gmv{\vec{\gamma} }
\def\alv{\vec{\alpha} }
\def\muv{\vec{\mu} }
\def\muh{\hat{\mu} }
\def\xbb{\overline{\overline{x}} }
\def\ybb{\overline{\overline{y}} }
\def\xb{\overline{x} }
\def\yu{\underline y\,}
\def\Yu{\underline Y\,}
\def\Xb{{\overline{{\bf{X}}}}}
\def\Ytu{\underline \tilde Y\,}
\def\Xtuu{{\bf \tilde X}\,}
\def\PXo{{\bf P}_{X_0}\,}
\def\Omuu{{\bf \Omega}\, }
\def\Siuu{{\bf \Sigma}\, }
\def\SiuuR{{\bf \Sigma}_{RC}\, }
\def\SiuuhR{{ \Siuuhat}_{RC}\, }
\def\Au{\underline A}
\def\alu{\underline\alpha}
\def\thu{\underline{\theta}}
\def\sgb{\overline{\sigma}}
\def\aluhat{ \underline{\hat\alpha} }
\def\beu{\underline\beta}
\def\Auu{{\bf A}\,}
\def\Lauu{{\bf \Lambda}}
\def\yb{\overline{y} }
\def\hb{\overline{h} }
\def\nb{\overline{n} }
\def\Bph{{\hat{B}_p}}
\def\ev{\vec{e}}
\def\xv{\vec{x}}
\def\wv{\vec{w}}
\def\ssgh{{\hat \sigma}^2}
\def\sstg{{\hat \sigma}_{2stg}}
\def\sgstg{{\hat \sigma}_{2stg}}
\def\ssg{\sigma^2}
\def\ybv{\vec{\overline{y}}}
\def\xbv{\vec{\overline{x}}}
\def\xmv{\vec{\overline{x}}}
\def\bvh{\vec{\hat{b}}}
\def\btv{\vec{ \beta}}
\def\btvh{\hat{\vec{ \beta}}}
\def\btd{\vec{ \beta}_{dl}}
\def\bthR{\hat{\vec{ \beta}}_{RC}}
\def\btbv{\vec{\overline{ \beta}}}
\def\btbvh{\hat{\vec{\overline{ \beta}}}}
\def\bvb{\vec{\overline{ \beta}}}
\def\xv{\vec{x}}
\def\wv{\vec{w}}
\def\xmv{\vec{\overline{x}}}
\def\bvh{\vec{\hat{ b}}}
\def\bvb{\vec{\overline{ \beta}}}
\def\nl{\hfill\break}       
\def\np{\vfill\eject}       
\def\ns{\vskip 2pc\noindent}       
\def\nsect{\vskip 2pc\noindent}       
\def\nli{\hfill\break\noindent}       
\def\ni{\noindent}
\def\IR{I\kern-.255em R}
\def\Ze{Z_{eff} }
\def\Zeff{Z_{eff} }
\def\tE{\tau_{E} }
\def\nb{\overline{n} }
\def\eps{\epsilon }
\def\llt{\lambda_T }
\def\pbs{\overline{p} }
\def\pbonn{\overline{p_1} }
\def\ptwb{\overline{p_2} }
\def\yu{\underline y\,}
\def\Yu{\underline Y\,}
\def\Xuu{{\bf X}\,}
\def\Siuu{{\bf \Sigma}\, }
\def\Siuuhat{{\bf {\hat\Sigma}}\, }
\def\XSX{{\Xuu^t \Siuuhat^{-1} \Xuu}\, }
\def\Dlh{{\bf {\hat\Delta}}\, }
\def\Suu{{\bf S}\, }
\def\muu{\underline{\mu}\, }
\def\muuhat{\underline{\hat\mu}\, }
\def\muuhathat{\underline{\hat{\hat\mu}}\, }
\def\eu{\underline e\,}
\def\wdr{w_{d,r}}
\def\Wuu{{\bf W}}
\def\Au{\underline A}
\def\Bu{\underline B\,}
\def\Eu{\underline E\,}
\def\alu{\underline\alpha}
\def\aluz{{\underline\alpha}^{(0)}}
\def\aluhat{ \underline{\hat\alpha} }
\def\eps{ \epsilon}
\def\bv{\vec \beta}
\def\beu{\underline\beta}
\def\cur{{\underline c}_r}
\def\curt{{\underline c}_r^t}
\def\euu{{\bf e}\,}
\def\Auu{{\bf A}\,}
\def\Buu{{\bf B}\,}
\def\Cuu{{\bf C}\,}
\def\Euu{{\bf E}\,}
\def\Iuu{{\bf I}\,}
\def\Luu{{\bf L}\,}
\def\Muu{{\bf M}\,}
\def\Puu{{\bf P}\,}
\def\Quu{{\bf Q}\,}
\def\Lauu{{\bf \Lambda}}
\def\Dl{{\bf \Delta}}
\def\Vuu{{\bf V}\,}
\def\Vuuhat{{\bf{\hat V}}\,}

\begin{center}
{\bf ON DIMENSIONALLY CORRECT POWER LAW SCALING
EXPRESSIONS FOR
L MODE CONFINEMENT}\\

K.S. Riedel \\
Courant Institute of Mathematical Sciences \\
New York University \\
New York, New York 10012
\end{center}

\begin{abstract}
Confinement scalings of divertor and radiofrequency heated
discharges are shown to differ significantly from the standard neutral
beam heated limiter scaling.
The random coefficient two stage regression algorithm is applied
to a neutral beam heated limiter subset of
the ITER L mode database as well as a combined dataset.
We find a scaling similar to Goldston scaling for the NB limiter
dataset and a scaling similar to ITER89P for the combined dataset.
Various missing value algorithms are
examined for the missing $B_t$ scalings.
We assume that global confinement can be approximately described
a power law scaling. After the second stage, the constraint
of collisional Maxwell Vlasov similarity is tested and imposed.
When the constraint of collisional Maxwell Vlasov similarity is imposed,
the C.I.T. uncertainty is significantly reduced while the I.T.E.R.
uncertainty is slightly reduced.  
\end{abstract}
\np
Global scaling expressions are widely used to analyse, interpolate,
and extrapolate tokamak performance$^{1-13}$. Initial efforts concentrated
on applying simple ordinary least squares regression using the dimensional,
``engineering", variables. Recent research has concentrated on dimensionless
scalings and on incorporating the tokamak to tokamak variation into
the regression analysis. In this article, we apply
the random coefficient (R.C.) two step regression procedure of Refs. [4,5]
while requiring the resulting expression to be dimensionless.

In Ref. [5], we showed with S. Kaye that
tokamak to tokamak variation accounts for over $90\%$
of the total variance of the scalings.
To model this tokamak to tokamak
variation, we treat the scaling differences between devices as random
variables. This probabilistic treatment is correct when the
tokamak to tokamak differences are due to many small factors.
If, however, this tokamak to tokamak variation is attributable to one or more
important factors such as wall material or limiter/divertor configuration,
statistics is of little help in analyzing confinement.

We begin by estimating a dimensional scaling expression
using the random coefficient two step regression procedure of Refs. [4,5].
The precise algorithm is discussed in detail in Refs. [4,5].
We briefly summarise the method.

First, for each tokamak, a scaling and covariance is estimated in
$I_p, \ B_t, \ \nb $ and $P$. We calculate the empirical mean
and covariance of these within tokamak scalings using the Swamy
random coefficient weighting procedure. Second, the mean confinement
time of each tokamak is corrected for the within tokamak scalings.
The scalings with $R/a,\ \kappa $ and $R$ are estimated by
regressing the corrected mean energy times of the tokamaks.
The error, $\SiuuR$ in our estimate, $\bthR$, of the scaling vector is
given by Eq. (18a) of Ref. 4.

\ns
{\bf II. Improved Neutral Beam Limiter Confinement Scaling}

In the random coefficient model, we assume that all confinement
difference are random and not systematical. Since our
initial statistical analysis, in collaboration with S. Kaye,
we have realised that the variation in the within tokamak
scalings is significantly less with only neutral beam heated (N.B.)
limiter discharges in comparison to divertor discharges and
radiofrequency heated (R.F.) discharges.
As noted in Refs. [11,12], the overall tendency is that
divertor and radiofrequency heated discharges tend to
have somewhat stronger density scalings and somewhat weaker
current dependencies. Table 1 gives the database summaries
for the various types of discharges.
\footnote{Throughout this article, we describe the plasma
current , $I_p$,
in units of MAmperes, the toroidal magnetic field, $B_t$ in
units of Teslas, the total heating power $P$ in
MWatts, and the line averaged
plasma density in $10^{19}$ particles per cubic meter.}
In Table 2, we present the within
tokamak scalings for various limiter/divertor configurations and
heating types. The modified lower x point configuration of
JT-60 is denoted by JTLX and is treated as a separate device.

A second feature is that the data scatter
and the variation in scalings tends to be significantly larger
when divertor discharges and R.F. discharges from different
tokamaks are compared. This larger variability may be
an artifact of that the ITER L mode database contains relatively
few R.F. and divertor discharges. Furthermore, the ITER database
was developed when neutral beam technology had already matured
while  R.F. heating was still largely under scientific
study.

 The R.F. and divertor data consists almost exclusively
of Japanese and Soviet tokamaks. We note that the N.B limiter
scalings of JT-60 and JFT-2M differ significantly from the
typical N.B. limiter scaling. In fact, the difference between
the R.F. and divertor scalings and the N.B. limiter scalings
on JT-60 and JFT-2M appear to be less than the departure of
the JT-60 and JFT-2M N.B. limiter scalings from the norm.
Nevertheless, the JET divertor scaling collaborates the
observed tendency of stronger density and weaker current scalings.

From Table 1, it is clear that the present database
has insufficient divertor or R.F. data to perform a
separate analysis. Thus we perform an analysis of only
N.B. limiter data and an analysis of a combined dataset containing
divertor and R.F. discharges as well. The results of the
combined analysis depend on the existing mixture of datapoints
and will systematically vary as more R.F. or divertor discharges
are added.

We begin with an ordinary least square regression analysis of
the 1346 datapoint combined database:

$\tau_E M^{-1/2} =$
$$.0351
\left( {R/a \over 3.62} \right)^{-.42}
\left( {R \over 1.83} \right)^{1.60}
\left( {\kappa \over 1.17} \right)^{.59}
\left( {I_p \over .606} \right)^{.79} \left(
{B_t \over 2.217} \right)^{.13} \left(
{\nb \over 3.947} \right)^{.10} \left(
{P \over 3.593} \right)^{-.47}
\ .
\eqno (1)$$

In contrast, our restricted N.B limiter database has a
ordinary least squares scaling of

$\tau_E M^{-1/2} =$
$$.0383
\left( {R/a \over 3.34} \right)^{.39}
\left( {R \over 1.84} \right)^{1.38}
\left( {\kappa \over 1.134} \right)^{.61}
\left( {I_p \over .7005} \right)^{.98} \left(
{B_t \over 2.138} \right)^{.13} \left(
{\nb \over 4.58} \right)^{.00} \left(
{P \over 4.09} \right)^{-.55}
\ .
\eqno (2)$$

We note that the aspect ratio and size scalings are significantly
modified when the combined database is used. In particular,
it is difficult to believe that there is such a
strong difference in aspect ratio scalings for different discharge
types. We note that {\it
the major difference between the ITER89P scaling
and the Goldston or Riedel-Kaye scalings is that the latter
scalings are derived almost exclusively on neutral beam limiter
data while the ITER89P scaling attempts to describe the combined
dataset.}

Our neutral beam heated limiter discharge analysis is based on a
705 datapoint subset of ITER L mode database. Our dataset consists of
one small tokamak, ISX-B,
three moderate size tokamaks, ASDEX, DIII and PDX and three large
tokamaks, JET, JT-60 and TFTR.
We restrict our analysis to discharges
with $q_{shaf} \le 6$. We assume that the isotope enhancement factor
is $M^{1/2}$.
The $\kappa$ scaling is treated as a between tokamak variable,
instead of being determined by the $\kappa$ scalings in DIII and ISXB.

Since our previous L mode scaling analysis$^5$, we have added several
new  restrictions on the data selection procedure. First we restrict
to limiter discharges 
since divertor discharges often
have somewhat different scaling characteristics. With this restriction,
the number of JET discharges is reduced from 149 to 93 and the number
of JT-60 discharges is reduced from 199 to 172. The JET confinement
scaling is thereby modified from
$\tau_E \sim  I_p^{.81\pm.06} B_t ^{.56\pm.05}
\nb ^{.08\pm.04} P^{-.65\pm.02} $ to
$\tau_E \sim  I_p^{.90\pm.07} B_t ^{.44\pm.05}
 \nb ^{-.05\pm.04} P^{-.57\pm.03} $.
Similarly, the JT-60 scaling is modified from
$\tau_E \sim  I_p^{.68\pm.02}   \nb ^{.18\pm.02} P^{-.66\pm.03} $ to
$\tau_E \sim  I_p^{.80\pm.02}   \nb ^{.08\pm.02} P^{-.57\pm.02} $.


We have eliminated five low power or low density ASDEX datapoints, 
which modifies
the ASDEX power scaling from  $\nb^{-.27\pm.13}  P^{-.26 \pm .06}$.
to $\nb^{-.17\pm.09} P^{-.32 \pm .14}$. Finally,
we have eliminated eight low current ISXB datapoints, which modifies
the ISXB  scaling from $I_p^{1.42 \pm .07}$ to
$I_p^{1.30 \pm .12}$.

A comparison of Table 2a with Table 2 of Ref. 5 shows that the
within tokamak scalings vary significantly less in the
new restricted dataset.
Thus these restrictions result in a more uniform dataset which better
characterises normal NBI limiter discharges.

In these low power or current discharges, the parametric dependencies
are somewhat different than the typical L mode scaling. This indicates
that the power law approximation to the functional form of $\tau_E$
is beginning to break down. In reality, plasma transport is a complicated
nonlinear function of many parameters. A power law ansatz corresponds
to a selfsimilar scaling of the dominant loss mechanisns. 

As discussed in Ref. 9, the log linear form can be viewed as a Taylor
series expansion of the actual functional form of $\tau_E$ about the
center of mass of the database. It should come as no surprise that
the L mode scaling changes and then breaks down as the extrapolation 
exceeds the original domain of validity. Many tokamaks have observed 
saturation in the current scaling below $q_a$ of three. In fact,
it is surprising that the L mode power law scaling works so well
over the ``standard" parameter subdomain of auxilary heated
tokamaks.

When the log linear functional form is applied in too large a
parameter subdomain, the statistical analysis treats the 
systematic errors from the unresolved functional form as
random errors.
The precise domain of application of L mode power law scalings is a 
crucial and as yet unresolved area in confinement physics.

\ns
{\bf III. MISSING VALUE PROCEDURES}

Since three (five for the combined database)
of the  tokamaks, ASDEX, JT-60, TFTR, (T-10 and JFT2M for the combined
database,)
 have no $B_t$ variation, their $B_t$ scalings can be inferred
by several different missing value algorithms.
The reason for using a missing value procedure is to produce
a complete second stage dataset.
Since our results depend somewhat on the choice of missing value
algorithm, we examine four alternatives.

The first possible missing value algorithm is to simply set
the $B_t$ dependence of $\tE$ in ASDEX, JT-60, TFTR, (T-10, and JFT2M)
equal to zero. Since many experiments observe no or an extremely
weak $B_t$ dependence, this simple approach is a
good first approximation. {\it The actual Riedel-Kaye scaling of Ref.5
used this algorithm.} A second missing value algorithm is to replace
the missing $B_t$ dependencies by the mean value of the $B_t$ scalings
in DIII, JET, ISXB, PDX (and JT60LX).

In Ref. 5, it was noted that the sum of the $B_t$ and $I_p$ scalings
tended to be a constant. Therefore, a third missing value algorithm
was proposed but not implemented in Ref. 5. This third algorithm
consists of fitting a straight line through the $B_t$ and $I_p$
scalings.
The four(or five)
tokamaks with $B_t$ scans were used to determine the free parameters,
$c_0$ and $c_1$ in
$\beta_{B_t} = c_0 + c_1 \beta_{I_p}$,
where $\beta_{B_t}$ is the scaling with $B_t$ and $\beta_{I_t}$ is the
scaling with $I_p$.
Then the ASDEX, JT-60, and TFTR $B_t$ scalings were inferred
(as well as T-10 and JFT2M for the combined database).
All three of these procedures are interpretive in the sense that
they substitute semiempirical values for the missing values.

Finally, the "projection"  missing value algorithm consists of
using only the principal components of the within
tokamak scalings which are estimatable. This projection algorithm
requires essentially no apriori assumptions about the missing
scalings.  The projection missing value
algorithm has another advantage, it can be more easily applied to
cases where in one or more tokamaks, other scaling directions,
i.e. principal components, have not been varied sufficiently
to be determined.
The disadvantage to the projection method is that the
projected data may be unbalanced and therefore illconditioned.

Virtually every reasonable missing value procedure will
systematically lower the estimates of the variance, because we
are replacing the random component of the $B_t$ scaling with
a more deterministic procedure. Since roughly half of the tokamaks
have no $B_t$ variation, we may underestimate the $B_t$ variance
by a factor of two.

Table 3 summaries the N.B. Limiter
R.C. scalings for the various missing value
algorithms. The $B_t$ coefficient varies from $.06$ to $.20$,
with the strongest $B_t$ dependence occuring when $\beta_B$
is regressed against $\beta_I$.
Table 4 presents the same comparison for the combined heating
and magnetic configuration database.

A crude measure of the relative merits of
each of the missing value procedures can be obtained by comparing
the residual sum of squares in the second stage regression
on the corrected centers of mass of the various tokamaks.
This measure of goodness of fit is rather inaccurate, since
we are fitting four free parameters to seven datapoints.
A second, independent measure of the merits of each missing
value procedure
is the extent to which a scaling intrinsically satisfies
collisional Maxwell Vlasov similarity (see Sec. IV).

 In Table 3c, 4c,
the second column, $\sstg$,  is the R.M.S.E. for the second stage
regression on the mean confinement times, weighted by the square
root of the number of degrees of freedom.
The third and fourth columns
give the predicted energy confinement time and estimated
statistical uncertainties for I.T.E.R. and C.I.T..
The fifth column is the ratio of the squared dimensional component of
to its variance.

We find that the projection missing value procedure has the smallest
root mean squared errors (RMSE) relative to the other missing value
procedures. Thus the projection procedure shows no signs of
illconditioning.
The projection algorithm also satisfies  C.M.V. similarity
to a greater extent than the three interpretive missing value
algorithms.
Also, the projection procedure makes the weakest
assumptions on the relational dependencies of the $B_t$ scaling.
Therefore we prefer the projection algorithm to the three
"interpretive " missing value algorithms.
The projection missing value procedure yields the following
N.B. limiter scaling:

$\tau_E M^{-1/2} =$
$$.0381
\left( {R/a \over 3.34} \right)^{.28}
\left( {R \over 1.84} \right)^{1.22}
\left( {\kappa \over 1.134} \right)^{.55}
\left( {I_p \over .7005} \right)^{1.02} \left(
{B_t \over 2.138} \right)^{.14} \left(
{\nb \over 4.58} \right)^{.01} \left(
{P \over 4.09} \right)^{-.54}
\ .
\eqno (3)$$

\footnote{We present our scalings centered about the database
mean, thus the mean values of our database are apparent. Also
if the scaling coefficients are rounded, the overall constant 
in the centered formulation does not need to be adjusted.
The overall constant in the noncentered version should be corrected
to match the overall constant of the centered formulation.}

For the combined dataset, we find that the RMSE is the smallest for
the ``$B_t = 0$" missing value procedure. However, the resulting
scaling has a noticable dimensional component. The projection
procedure has a R.M.S.E. comparable to the ``$B_t$" mean scaling
procedure and an smaller dimensional component.
Thus we select the projection procedure again.
The corresponding R.C. regression for the combined dataset
yields:

$\tau_E M^{-1/2} =$
$$.0346
\left( {R/a \over 3.62} \right)^{-.36}
\left( {R \over 1.83} \right)^{1.55}
\left( {\kappa \over 1.17} \right)^{.63}
\left( {I_p \over .606} \right)^{.86} \left(
{B_t \over 2.217} \right)^{.18} \left(
{\nb \over 3.947} \right)^{.15} \left(
{P \over 3.593} \right)^{-.525}
\ .
\eqno (4)$$

We denote the vector of scaling coefficients by $\bthR$.
Table 5 gives the $8x8$ covariance matrices,
$\SiuuR$, for our two-step regression vector
$\bthR$ as derived in eqn. 18 of [4].
 In evaluating $\SiuuR$,
we include the small discharge to discharge variation term which
was neglected in Ref. [5].

To evaluate the statistical uncertainty in the predicted
energy confinement for a given set of parameters,
we transform the tokamak's parameters
to the centered logarithmic variables, $\vec{x}^t$,
and take the interproduct with the covariance matrix of Table 1.
The centered $\xv^t$ variable is

$$
\left(
( \ln R - .609) \ ,
\left( \ln {R \over a} - 1.206 \right) \ ,
( \ln \kappa - .126) \ ,
1 \ , \right.
$$
$$
\left. ( \ln I_p  + .356) \ , ( \ln B_t - .760) \ ,
( \ln \nb - 1.522) \ ,
( \ln P - 1.409) \right) \ ,
$$
for the N.B. limiter dataset and

$$
\left(
( \ln R - .605) \ ,
\left( \ln {R \over a} - 1.29 \right) \ ,
( \ln \kappa - .154) \ ,
1 \ , \right.
$$
$$
\left. ( \ln I_p + .501) \ , ( \ln B_t - .796) \ ,
( \ln \nb - 1.373) \ ,
( \ln P - 1.279) \right) \ ,
$$
for the combined dataset.
The fourth index corresponds to the absolute constant in the
scaling law.

In Tables 3c,4c, $\sgstg$, the second stage R.M.S.E., corrected for 
the number of degrees of freedom, is around three to five percent for
the N.B. limiter dataset and between six and eight percent for 
the combined dataset. Thus the fit on the combined dataset is
significantly worse. The R.C. variance in our model consists
of two terms. First, the variance of the absolute constant
is precisely equal to $\sgstg$ divided by the number of tokamaks.
The second term is the variance of the within tokamak scalings.
For both datasets, the within tokamak scaling variance dominates
the total variance estimate. Thus the principal reason why the 
combined regression has a larger variance is the larger within
tokamak scaling differences and not the increase in R.M.S.E..

For ITER, we assume the following parameter value:
$M = 2.5$, $a = 2.15 m$, $R = 6.0m$, $\kappa = 2.0$, $I_p = 22MA$,
$B_t = 5T$, $\nb = 13.8 \times 10^{19}$, $P_{\rm tot} = 150 MW$.
The resulting predicted confinement times is
2.27 sec with an uncertainty
factor of 21\% for the combined dataset scaling and
1.82 sec with an uncertainty
factor of 14\% for the N.B. limiter  dataset scaling.
Clearly the predicted confinement times only weakly depend 
on the choice of missing value
 procedure for the $B_t$ scaling.

For CIT, we use the following parameter values: $M = 2.5$,
$a = .65 m$, $R = 2.1m$, $\kappa = 2.0$, $I_p = 11MA$,
$B_t = 10T$, $\nb = 50 \times 10^{19}$, $P_{\rm tot} = 100 MW$.
We predict a CIT L mode confinement time of 392 msec with an uncertainty
factor of 22 \% for the combined dataset scaling
and a L mode confinement time of 364 msec with an uncertainty
factor of 24 \% for the N.B. limiter scaling.
The predicted confinement of C.I.T. varies about $20\%$ depending
on the choice of missing value procedure. This variation is still
within the error bars. Nevertheless, the differences indicate
the sensitivity of extrapolation to high field devices while the
present database has little $B_t$ variation at constant size.

Since the R.M.S.E. of the combined dataset is nearly four times
larger than the N.B. limiter dataset, it is difficult to understand
why the C.I.T. error estimate is smaller. 
the C.I.T. uncertainty estimate depends strongly on the $B_t$ scaling
variance. As noted earlier, the missing value procedures systematically
underestimate the $B_t$ scaling variance. 
The combined dataset has a larger ratio (5/10) of tokamaks with no
$B_t$ variance than the N.B. dataset (3/7). Thus the underestimate may be larger
for the combined dataset.
 This may partially explain the
slightly smaller C.I.T. uncertainty estimate.
A second reason is that the error in our R.C. estimates is sufficiently
large that the errorbars of the two estimates overlap. 


In the second stage regression, we are fitting the corrected
mean tokamak
confinement as a function of four free parameters.
Usually the use of four parameters to fit seven datapoints
( or even ten datapoint for the combined dataset) would be
considered overfitting.
The smallest principal component of the second stage regression
accounts for only about one percent of the total variance for
the N.B. dataset and  .02-.03 for the combined dataset
Thus we initially believed that we could eliminate the smallest
principal component. 
However dropping the last component raises the RMSE by a factor of up
to ten. The systematic errors are usually on the order of five to
ten percent. Thus changes in the goodness of fit from $2-3.5\%$ to
$10\%$ are significant. We therefore keep all principal components
in the second stage regression.
\ns
{\bf IV. COLLISIONAL MAXWELL VLASOV CONSTRAINT}

 When a particular class of physical phenomena are responsible
for anomalous transport, the resulting scaling expression should possess
the same similarity transformations as the underlying physics instability.
We consider the case where the turbulent transport is well described by
the  collisional Maxwell Vlasov (C.M.V.) system
and the ratio of the Debeye length to all other scale
lengths is infinitesimally small.
This system is completely prescribed by three dimensionless
variables$^{14,15}$:

$$ \beta \equiv \nb T_i/B_t^2 \ , \rho_i* \equiv (M T_i)^{1/2} / RB_t \ ,
\nu_i* \equiv R \nb q /T_i^2
\eqno (5)$$
\noindent
together with the four naturally dimensionless variables: $\kappa, R/a, q_{cyl}$
and $M$. Thus $\tE \Omega_i$ is completely describable as a function of
these seven variables.
We eliminate the temperature dependence in eqn(5)
in favor of $\tE$ using $\tE P = <n T> Vol$.

To determine dimensionless scaling
expressions, we assume that the $\tE \Omega_i$ can be modelled as
a {\it log linear function of the dimensionless variables.}
For log linear functions of the C.M.V. variables, $B_t \tE$
is a function of only
$ v_1 = ln \left( {P  \over R^3 B_t^3} \right),\
v_2 = ln \left( {M P  \over \nb R^5 B_t^3} \right),\
v_3 = ln \left( {P  \over R^{7/2} \nb^{3/2} B_t} \right),$
along with the four naturally dimensionless variables: $\kappa, R/a, q_{cyl}$
and $M$.

We treat this hypothesised collisional Maxwell Vlasov similarity as a constrained
regression. As shown in Ref. [4], for power law scalings, this
linear constraint reduces to

$$ \gmv \cdot \btv = - \ev_B \cdot \gmv = - \gamma_B
 \eqno (6)$$
\noindent
where $\gmv$ is orthogonal to the 7 dimensional space spanned by
the dimensionless $\alv_k$. In our case,

$$\gamma_{R/a} = 0, \ \gamma_{\kappa} = 0, \
\gamma_R = 1, \ \gamma_{const} = 0, \
\gamma_{Ip} = -1/4, \  \gamma_{Btor} = -5/4, \
\gamma_P = -3/4, \ \gamma_{\nb} =-2  .
 \eqno (7)$$

We note that generally the size scaling is indeterminable
within a given tokamak. Thus only by comparing a number of tokamaks can
we determine a size scaling and therefore examine the C.M.V. constraint.
When the major and minor radius are varied within a single device,
the distance to the wall is also varied. Thus it is difficult to
determine if the size scaling experiments in T.F.T.R. are heavily
influenced by changes in $\Zeff$ as the shape is varied.

Therefore we determine a C.M.V. constrained scaling within the
multiple tokamak R.C. analysis.
We denote the unit vector in the $\gmv$ direction by $\gh$.
To determine $\btd$, we minimise the restricted least squares
functional:
$$
\min_{\btd} ( \btd - \bthR )^t \cd  \Siuu_{RC}^{-1}
\cd ( \btd - \bthR )
 + \lambda ( \gmv \cdot \btd + \gamma_B   )
\ . \eqno (8)
$$
\noindent

The solution is
$$
\btd =  \bthR + \lambda \SiuuR \cd \gmv
\eqno (9) $$
\noindent
where $ \lambda =  -
( \gmv \cdot \bthR + \gamma_B  )  /
\left( \gmv^t \cd \SiuuR \cd  \gmv \right). $

This dimensionless scaling expression minimises the difference
between the generalised least squares estimator of Eqs. 3,4 and any
dimensionless scaling expression as measured by the $\SiuuR^{-1}$
metric. For any given metric, $\Siuu_{arb}^{-1}$, Eq. 9 yields the
corresponding minimising dimensionless expression, where $\SiuuR$
is replaced by $\Siuu_{arb}$. {\it The common practice of arbitrarily
adjusting the coefficients of a dimensional scaling to make it
dimensionless, results in suboptimal scalings which can differ
significantly from the closest dimensionless scaling.}

To test for C.M.V. similarity in log linear scalings, we assume
that the statistical model of Refs. [4,5] is correct, i.e.
$\tE$ has a log linear scaling with the tokamak to tokamak
variation being given by the R.C. model of Ref. [4]. We
now test if within this model, we can impose the additional
constraint on $\btv$ given by Eq. 6. The expected deviation
from the hypothesised dimensionless scaling is

$$ Exp \left[( \btd - \bthR )^t \cd  \Siuu_{RC}^{-1}
\cd ( \btd - \bthR ) \right] = \lambda^2 \gmv^t \cd \SiuuR \cd  \gmv
= ( \gmv \cdot \bthR + \gamma_B  )^2 / \gmv^t \cd \SiuuR \cd  \gmv
\eqno (10)$$

$\bthR$ and $\SiuuR$ are basically the empirical mean and variance
of the scalings of seven different tokamaks and therefore can be
modeled with a $T^2$ distribution$^{16}$. Since we are interested
in a single fixed component, $\gmv \cd \bthR$, the
relevant test statistic is $T^2 \equiv
| \gmv \cdot \bthR + \gamma_B |^2 / {\gmv^t \cd \SiuuR \cd  \gmv}$.
The $T^2$ statistic for this component has a $F(1,6)$ distribution,
($F(1,9)$ for the combined dataset.) The $F(1,n)$ distribution
is the generalisation of the $\chi$ distribution to the case of
an empirically determined variance, i.e. the Student T distribution.
If ${\gmv^t \cd \SiuuR \cd  \gmv}$ were known and not estimated,
$T^2=1$ would correspond to one standard deviation and
$T^2=4$ would correspond to two standard deviations.
The  $50 \% \ $ confidence level (corresponding to the halfwidth)
for the $F(1,6)$ distribution is  $T^2 = 0.515$.
and for the $F(1,9)$ distribution is  $T^2 = 0.494$.
The  $95 \% \ $ confidence level 
for the $F(1,6)$ distribution is  $T^2 = 5.99$.
and for the $F(1,9)$ distribution is  $T^2 = 5.12$.

For the L mode dataset, with the projection missing value
 procedure,
the test statistic, $T^2 \equiv
| \gmv \cdot \bthR + \gamma_B |^2/ {\gmv^t \cd \SiuuhR \cd  \gmv}
= 0.168$ for the NB limiter dataset and 0.044 for the combined dataset.
{\it These $T^2$ values are so small that we can not only set the
dimensional projection equal to zero, but also eliminate the
R.C. variance in the dimensional direction from our uncertainty
estimates.}

The use of the $T^2$ distribution is only strictly valid for within
scaling vectors. The precise probability distribution of between
scaling vectors is almost indeterminable.
Since we are not interested
in the tail of the probability distribution,
this influences our results only weakly.


Of course, if all the edge physics, deposition physics and radiative
losses were accurately modeled the hypothesised dependence on
collisional Maxwell Vlasov variables should be trivially true. Nevertheless,
the power law form in C.M.V. variables would still be a crude approximation.
We can interpret the "extra" dimensional variable as an auxilary moment
of the input variables. Thus we could view our statistical hypothesis
as an attempt to eliminate the use of this auxilary moment in our
modeling of confinement.

The constrained scaling vector given
by Eq. (6):

$\tau_E M^{-1/2} =$
$$.0383
\left( {R/a \over 3.34} \right)^{.270}
\left( {R \over 1.84} \right)^{1.277}
\left( {\kappa \over 1.134} \right)^{.548}
\left( {I_p \over .7005} \right)^{1.009} \left(
{B_t \over 2.138} \right)^{.133} \left(
{\nb \over 4.58} \right)^{.0096} \left(
{P \over 4.09} \right)^{-.548}
\ .
\eqno (11)$$
We give the scaling coefficients to three digits accuracy,
not because of precision, but to reduce the extent which rounding
error induces a violation of C.M.V. similarity.
The constrained scaling for the combined database is
$\tau_E M^{-1/2} =$
$$.0346
\left( {R/a \over 3.62} \right)^{-.367}
\left( {R \over 1.83} \right)^{1.575}
\left( {\kappa \over 1.17} \right)^{.629}
\left( {I_p \over .606} \right)^{.851} \left(
{B_t \over 2.217} \right)^{.171} \left(
{\nb \over 3.947} \right)^{.146} \left(
{P \over 3.593} \right)^{-.525}
\ .
\eqno (12)$$

The constrained, combined dataset
scaling of Eq. 12 yields
a predicted I.T.E.R. confinement time of 2.27 sec $\pm 21 \%$
and a predicted C.I.T. confinement time of .383 sec $\pm 17 \%$

If the dimensionless scaling expression is accepted, the variance,
$\Siuu_{dl}$, of the estimate, $\btd$, is the projection of $\SiuuR$
onto the dimensionless subspace, i.e.
$\Siuu_{dl}= \SiuuR - \SiuuR \gh \gh^t \SiuuR /(\gh^t \SiuuR  \gh)$.
If the dimensionless scaling expression is accepted, the estimated uncertainty
arising from the constrained random coefficient model with the projection
procedure is reduced to 
 $ 15-20 \% \ $ percent for both I.T.E.R. and C.I.T..  
The unaccounted for uncertainties are discussed in Ref. [5].

We note that the C.M.V. constraint reduces the estimated C.I.T. uncertainty
significantly more than the estimated I.T.E.R. uncertainty.
We conjecture that this arises for the following reason. Both I.T.E.R.
and C.I.T. parameters have been chosen largely from physics considerations.
Thus, in some sense, I.T.E.R. and C.I.T. have minimised the dimensional
component of the extrapolation subject to engineering constraints. However
since the present database lacks high field tokamaks, the extrapolation
to C.I.T. entails a larger dimensional extrapolation.
\ns
\noindent
{\bf V. DISCUSSION}

Global scaling expressions, in particular, the Goldston-Aachen scaling$^1$,
have been successful in predicting the energy confinement in the
present generation of large tokamaks.
Recently, a partial consensus$^{12}$ has emerged that confinement data
could best be fitted and extrapolated using the ITER89P scaling$^{11}$
or a Goldston-like scaling. We note that both the Goldston and
Riedel-Kaye scalings were derived and optimised for N.B. discharges.
The ITER89P scaling is based on a larger class of discharges.

As discussed earlier, when restricted to a single class of discharges,
the R.C. model becomes a statistically justifiable, in some senses
optimal, approach. However the most important use of scaling
expressions is to extrapolate confinement to next generation devices.
Since future ignition tokamaks will be divertor devices with
$\alpha$ particle and R.F. heating, it only seems reasonable
to include the R.F. and divertor discharges in the database
when extrapolating to reactor relevant plasmas. 

However, since the combined scaling is a weighted average
over a diverse set of operating conditions, it probably reflects
to many different transport processes. Thus Eq. 12 should not
be compared with specific theortical transport models.
The N.B. limiter scalings tend to be more uniform and may
be of use for physical understanding. We note that although
the N.B. limiter scaling is nearly uniformly observed, it
probably is strongly influenced by power deposition and
edge physics effects.
Eq. 11 is also probably more accurate in fitting the standard
N.B. limiter plasmas.
Much of the nonN.B. limiter data, on which the combined scaling is
based, has a mixed reputation. Thus supporters of scalings similar 
to Eq. 11 can reasonably argue that the combined dataset is not 
sufficiently reliable to modify the relatively solid N.B. scalings.
Assuming the additional data is noisier, it is unclear whether
the inclusion of the R.F. and divertor data results in more
accurate predictions for C.I.T. and I.T.E.R. .

An abstract summary of the situation is as follows. The standard
N.B. limiter database subset is probably of higher quality and
may differ systematically from the R.F. divertor discharges.
If there were no systematic differences and the relative variances
of the N.B. and R.F. data are known, the most accurate statistical
estimate of the scaling is given by weighting each tokamak 
inversely proportional to its variance.

Since we do not know the relative variances of   
the data from each tokamak, we have weighted all tokamaks in our
analysis equally. If the variance of the R.F. divertor data were 
considerably larger than the N.B. data, the equal tokamak weighting
incorrectly weights the tokamaks and may even have a higher variance 
than the analysis which excludes this additional data!

In reality, the major danger is systematic and not random errors.
The systematic differences occur from both different edge and
heating physics as well as possible poorer operational conditions. 
One approach to the systematic physics differences would be
to estimate the R.F. divertor scaling using the N.B. limiter 
scaling and estimated variance as a Bayesian prior.

Our analysis has not considered systematic differences and the
combined analysis weighted all tokamaks equally. In reality,
the consensus of the confinement community appears to be that
the N.B. limiter discharges in the database should be weighted
more heavily than the preliminary R.F. divertor data. This prior
information can be accomadated by assigning each tokamak in the
database an appriori variance $\Dl_k$ proportional to the estimated
random coefficient matrix, $\Dl$, i.e. $\Dl_k \equiv \alpha_k \Dl$ 
where $\alpha_k$ is given. 

A simple alternative to this reweighted random coefficient regression
is to average the N.B. Limiter scaling of Eq. 11 with the combined
scaling of Eq. 12. {\it We recommend a simple geometric average of the 
two scalings 
be used for extrapolating to future large scale devices.}
The geometric average of the two scaling is the arithmatic average
on the logarithmic scale
and corresponds to weighting the neutral beam limiter discharges
roughly a factor of two more than the R.F. divertor discharges.
This geometric average of two constrained log linear scalings
will automatically satisfy  collisional Maxwell Vlasov similarity.
This compromise scaling has the parametric representation,

$$\tau_E^{comp} =
.03686 M^{1/2} ({R/a})^{-.049} {R }^{1.426}
{\kappa}^{.588}{I_p}^{.930} {B_t}^{.152}
{\nb}^{.078} {P}^{-.537}
\ .
\eqno (13)$$

This geometric average of the two scalings also has the advantage
that it lies midway between the Goldston and ITER89P scalings which
represent the consensus of the confinement community.
Since the two scalings are not independent and in fact use roughly
the same data the standard formula for the variance of two independent
estimates of a predicted value is not applicable! We therefore
suggest that the variance of the averaged scaling for any predicted
value such as I.T.E.R. or C.I.T.
be approximated by the simple average of the variances 
for the two predictions.

The major practical difference between the Goldston and ITER89P
scalings or the N.B. limiter scaling of Eq. 11 and the combined
scaling of Eq. 12 is the exponent on the aspect ratio scaling.
The additional R.F. divertor data is at a higher aspect ratio
than most of the N.B. limiter data. {\it The additional 
data decreases the aspect ratio scaling because the new data
on average attains lower values of confinement than predicted
by the N.B. limiter scaling with a strong favorable aspect  
ratio scaling.}
The geometric average of the two scalings, Eqs. 11 and 12,
will have virtually no aspect ratio scaling.

The Riedel-Kaye scaling of Ref. [5]
is strikingly similar to the Goldston scaling. The Riedel-Kaye algorithm
differs from the original Kaye-Goldston$^3$ algorithm by not only
correctly weighting the scalings of the various tokamaks, but also
{\it by treating $\kappa$ as a between variable.} We suspect the
physics of small, unoptimised $\kappa$ variations in a single tokamak
is different than large variations in different devices. We speculate
that the old Kaye-Goldston algorithm, modified only
by treating $\kappa$ as a between variable, might yield a scaling
similar to Refs. [1,5].

Although power law scaling ansatz is a crude model which lacks a
physical basis, the traditional log linear scaling
incorporates all the major engineering variables with a similarity
type behavior. It has been our experience that the ITER L mode
database is too poorly structured to allow models with more
free parameters, such as offset linear scalings, to be reliably
fitted. In this article, we have examined whether a model with
fewer free parameters, the collisional Maxwell Vlasov model can be used
to model the data. To answer this question, we must accurately
model the errors associated with the scaling. Since we
have done this within the framework of our two stage RC model,
we now discuss how our choice of RC model effects the
analysis of the C.M.V. constraint.

By fitting a simple power law scaling to complex loss mechanisms,
we make systematic errors. This effects our statistical test of
collisional Maxwell Vlasov similarity in two ways. First, in our RC model,
these biased errors are interpreted as random errors, which
thereby increase our estimate of $\SiuuR$. Second, the extent to
which confinement violates collisional Maxwell Vlasov similarity may
be increased or decreased by the biased errors. Since our
statistical test is the ratio of these two terms, both of
which are biased upward, the overall tendency of the systematic
errors is difficult to assess.

As discussed in [4], the present data is insufficient to
compute the entire $\SiuuR$ matrix.
Thus the between and crosscovariance are specified
by our statistical model and not determined empirically.
Our choice of R.C. model is the
most nearly homoscedastic possible with
the empirically estimated within covariance, $\Dl$.
In this context, homoscedasticity means that the
variance of the random variable component of the scalings
has minimal parametric
dependencies. If the actual $\SiuuR$ differs significantly from
our nearly homoscedastic model, the real statistic for C.M.V.
similarity might be significantly different than that of
Sec. II.


We now discuss the advantages of our method of constraining the
variables to C.M.V. similarity relative to other approaches.
An extremely naive approach is to simply regress $B_t \tE$
versus the seven dimensionless variables.
Linear regression$^{9,10}$ postulates
that " no measurement errors occur in the $x$ variables".
Since the largest measurement errors
occur in $\tE$, the regresion needs to be formulated so that
most of the errors are in the dependent variable.
The naive regression using $\beta, \nu*$, and $\rho*$ is actually
more poorly conditioned since the dependent and independent variables
are both explicitly defined as powers of the poorly measured variable.

The next level of sophistication is to commonly termed the "power
formulation"$^{4,6-8}$ and involves explicitly using only three of
the four possible combinations of $R, \ B_t, \ \nb $ and $P$.
This procedure would be adequate if a simple least squares,
uncorrelated approach were sufficient. Unfortunately,
the tokamak to tokamak variation requires a two stage
R.C. regression.

In our present formulation, we have treated the individual
tokamak scalings as the basic observed quantity. This corresponds
to neglecting the within tokamak variation. We have also applied
the within tokamak test statistic for the constraint even though
the size scaling component of the constraint is in the between
tokamak direction. Our previous formulation of the C.M.V. constraint
treated the individual discharges as the basic observation and
used the R.C. matrix to determine a general $\Siuu$ matrix.
Although this earlier formulation is more general, and treated
both between and within variation, it is difficult to apply in
practice. The previous treatment required the use of $(\XSX)^{-1}$.
Since $\SiuuR$ is poorly conditioned, the estimate of $(\XSX)^{-1}$,
which is needed in Ref. 4, can be wildly inaccurate.

The concurrent work of Christiansen, et. al$^{14}$differs from this
work by applying ordinary least squares analysis to the L mode database.
Thus it treats the within tokamak errors while not considering the
between tokamak errors.

It is possible to perform a dimensionless two stage R.C. regression
using $q_{cyl}$ and the three dimensionless combinations of
$R, \ B_t, \ \nb $ and $P$ as within covariates. This amounts
to determining the size scaling as a within covariate and only
$R/a$ and $\kappa$ as between covariates. We strongly disfavor this
approach since {\it the size variation constitutes the largest
principal component and therefore the $R$ scaling is the
easiest to determine of the between scalings.}

In conclusion, we have treated the collisional Maxwell Vlasov similarity
ansatz for power law scalings as a constraint. Because
the random coefficient algorithm not only produces efficient
estimates of the parametric scalings but also a covariance
matrix for the errors in the scaling, we are able to test
this similarity ansatz. 
When the constraint of collisional Maxwell Vlasov similarity is imposed,
the C.I.T. uncertainty is significantly reduced while the I.T.E.R.
uncertainty is slightly reduced.

For future R.F. and alpha particle heated divertor 
experiments, we believe that the constrained, combined dataset
scaling of Eq. 12 represents the most reliable extrapolation
method. We find a predicted I.T.E.R. confinement time of 2.27 sec $\pm 21 \%$
and a predicted C.I.T. confinement time of .383 sec $\pm 17 \%$


{\it Acknowledgment}

The author thanks Geoff Cordey for discussions about dimensionless
scaling constraints.
Section II. was influenceed in part on Ref. 14, which was written in
collaboration with O. Kardaun, T. Takizuka and P. Yushmanov.
The author thanks F. Perkins, C. Bolton, R. Goldston, S. Kaye,
and K. Lackner
for repeatedly urging him to do dimensionless regressions.

This work was supported under U.S. Department of Energy Grant No.
DE-FG02-86ER53223.

\np
\begin{center}
{\bf REFERENCES}
\end{center}

\begin{enumerate}
\item Goldston, R., Plasma Physics and Controlled Fusion, {\bf 25}, 1984, p. 65.
\item Kaye, S.M., Goldston, R., Nucl. Fusion {\bf 25}, 1985, p. 65.
\item Kaye, S.M., Barnes C.W., Bell, M.G., et al,
Phys. of Fluids B, {\bf 2} p.2926, (1990).
\item  Riedel, K.S.,
Nuclear Fusion, {\bf 30}, No. 4, p. 755, (1990).
\item Riedel, K.S., Kaye S.M., { Nuclear Fusion},
{\bf 30}, No. 4, p. 731, (1990).
\item  Christiansen, J.P., Cordey, J.G., and Thomsen, K.,
Nuclear Fusion, {\bf 30}, No. 7, p. 1183, (1990).
\item Bickerton, R., Comments on Plasma Physics and COntrolled
Nuclear Fusion, Vol. 13, p.19, (1989), University of Texas 
Fusion Research Center Report No. 374, (1990).
\item Goldston, R., Proceedings of the 12th European Conference on
Plasma Physics and Controlled Nuclear Fusion, Vol. I, p.134,
Amsterdam, (1990)
\item Riedel, K.S.,
Comments in Plasma Physics and Controlled Fusion,
Vol. XII, No. 6, p.279, (1989).
\item Riedel, K.S.,
"Advanced Statistics for Tokamak Transport: Colinearity and
Tokamak to Tokamak Variation",
New York University Report MF-118, March, (1989),
National Technical Information Service document no. DOE/ER/53223-98.
\item Yushmanov, P., Takizuka, T., Riedel, K.S., Karadun, O., Cordey,
J., Kaye, S. and Post, D., Nucl. Fusion, {\bf 30}, p.?, (1990)
\item Doug Post and ITER Physics Team,
{\it ITER Physics Basis}, Section 3.1, "L mode confinement",
IAEA Vienna (1991) (to be published)
\item  Christiansen, J.P., Cordey, J.G., Kardaun, O.J.W.F.
and Thomsen, K.,
to be submitted to Nucl. Fusion.
\item Kadomtsev, B.B., Sov. J. Plasma Phys. Vol. 1 (1975) p.295
\item Hagan, W.F., Frieman, E.A., Phys. Fl.,
 Vol. 29, p.3635, (1986) .
\item Mardia, K.V., Kent, J.T. and Bibby, J.M., {\it Multivariate
Analysis}, Academic Press, London (1982).

\end{enumerate}
\end{document}